# Directional emission of nanoscale chiral sources modified by gap plasmons

Hai Lin[1], Te Wen[1], Jinglin Tang[1], Lulu Ye[1], Guanyu Zhang[1], Weidong Zhang[1], Ying Gu[1,2,3], Qihuang Gong[1,2,3], and Guowei Lu[1,2,3,*]

[1] *State Key Laboratory for Mesoscopic Physics, Frontiers Science Center for Nano-optoelectronics & Collaborative Innovation Center of Quantum Matter, School of Physics, Peking University, Beijing 100871, China.*
[2] *Collaborative Innovation Center of Extreme Optics, Shanxi University, Taiyuan, Shanxi 030006, China.*
[3] *Peking University Yangtze Delta Institute of Optoelectronics, Nantong 226010, Jiangsu, China.*
*Corresponding author: guowei.lu@pku.edu.cn*

**Efficient manipulation of the emission direction of a chiral nanoscale light source is significant for information transmission and on-chip information processing. Here, we propose a scheme to control the directionality of nanoscale chiral light sources based on gap plasmons. The gap plasmon mode formed by a gold nanorod and a silver nanowire realizes the highly directional emission of chiral light sources. Based on the optical spin-locked light propagation, the hybrid structure enables the directional coupling of chiral emission to achieve a contrast ratio of 99.5%. The emission direction can be manipulated by tailoring the configuration of the structure, such as the positions, aspect ratios, and orientation of the nanorod. Besides, a great local field enhancement exists for highly enhanced emission rates within the nanogap. This chiral nanoscale light source manipulation scheme provides a way for chiral valleytronics and integrated photonics.**

## 1. Introduction

Nanoscale light sources represented by emerging nanomaterials such as two-dimensional semiconductors, color centers, and quantum dots have showcased opportunities for nanophotonics and integrated optoelectronic devices 1-3. The chirality of light sources provides a new degree of freedom for integrated optical networks, bioimaging, and sensing 4-6. However, the poor emission directionality of nanoscale sources can affect the signal transmission and collection, thus limiting their practical application in optoelectronic integrated systems 7. In highly confined light fields of the subwavelength structure, transverse spin angular momentum (TSAM) can naturally appear with its handedness locked with the propagation direction of photonic modes 8-9. This phenomenon can be implemented to induce unidirectional transmission via circularly polarized dipoles coupled to the evanescent wave with the corresponding TSAM 8. The chiral coupling of photons and the emitter is non-reciprocal since the forward and backward propagating photons interact differently with the emitter 8, 10. The nonreciprocity in chiral coupling has led to various experiments on integrating non-reciprocal optical elements, including classical and quantum systems, such as low-loss isolators at the single-photon level 11, single-photon optical diodes 12, multiport circulators 13, all-optical switches 14, quantum gates 15, and chiral spin-entangled quantum networks 16-17. The transverse optical spin-dependent coupling in spin-path locked systems of chiral light sources has attracted lots of research 18-20. For example, the coupling between tungsten disulfide and silver nanowire can provide directional photoluminescence emission 21-22. Besides, unidirectional emission can be realized through dielectric waveguides such as dielectric nanowires, optical fibers, photonic crystal waveguides, and asymmetric metasurfaces 23-30. Recently, dielectric and nanostructure-free structures allow valley-selective exciton emission routing in 2D TMDCs (transition metal dichalcogenide) monolayer 31. Though highly directional transport can be achieved in spin-path locking systems with appropriate selection rules, these structures' coupling rate is low.

It is well known that localized surface plasmon resonance (LSPR) can enhance nanoscale light-matter interactions by concentrating electromagnetic fields in tiny volumes 32. V-shaped optical antennas, chiral double-rod antennas, and chiral metasurfaces have been demonstrated to control the emission direction of a nanoscale light source 33-35. LSPR systems can obtain optical spin-resolved properties through chiral interactions with a higher coupling rate but low emission directionality. Combining LSPR with a waveguide, the system has optical spin-resolved properties and a high coupling rate, which enable efficient directional emission of nanoscale light sources. In the Nanowire-Nanorod structure, the cavity modes with opposite parities enable the interference of odd and even modes, generating the asymmetric hybrid plasmon mode with strong near-field directionality 36. In recent work, a system based on the chiral coupling between emitters and gap plasmons composed of GaAs nanowire and silver nanoblock was proposed to obtain non-reciprocal photon transmission with a directionality of 91% 37. However, in the proposed scheme, the circularly polarized dipoles oscillate in the gap plane between the nanoblock and the nanowire. The assembly and precise positioning of the gap plasmon system based on in-plane chiral photon-emitter coupling are challenging and difficult to implement experimentally.

Here, we demonstrate a hybrid structure based on the vertical stacking of the gold nanorod and silver nanowire. The gap plasmon modes formed in the gap between the nanowire and nanorod are highly confined, providing localized transverse optical spin with a one-to-one relation between the handedness of optical spin and the propagation direction of a photonic mode. Nanoscale light sources are preferentially coupled to photonic modes, with TSAM corresponding to the chirality of light sources. Then, we investigated the influence of the nanogap size, the relative position, the aspect ratio, and the orientation of the nanorod on directional coupling. The emission is unidirectional when the hybrid structure is non-axisymmetric. Therefore, the directional emission of nanoscale light sources can be manipulated by controlling the structure's configuration. The simulation results show that the directional emission of chiral nanoscale light sources coupled with the nanowire can reach a directionality of 99.5%. Besides, the local field enhancement of up to ~1500 times at the gap between the gold nanorod and the silver nanowire can significantly increase the emission rate. The proposed hybrid structure has the characteristics of high stability and broad working bandwidth, which provides a more feasible solution for experimentally tuning the emission direction of chiral nanoscale light sources. The gap-plasmon system will be useful for on-chip non-reciprocal devices, such as optical isolators and circulators.

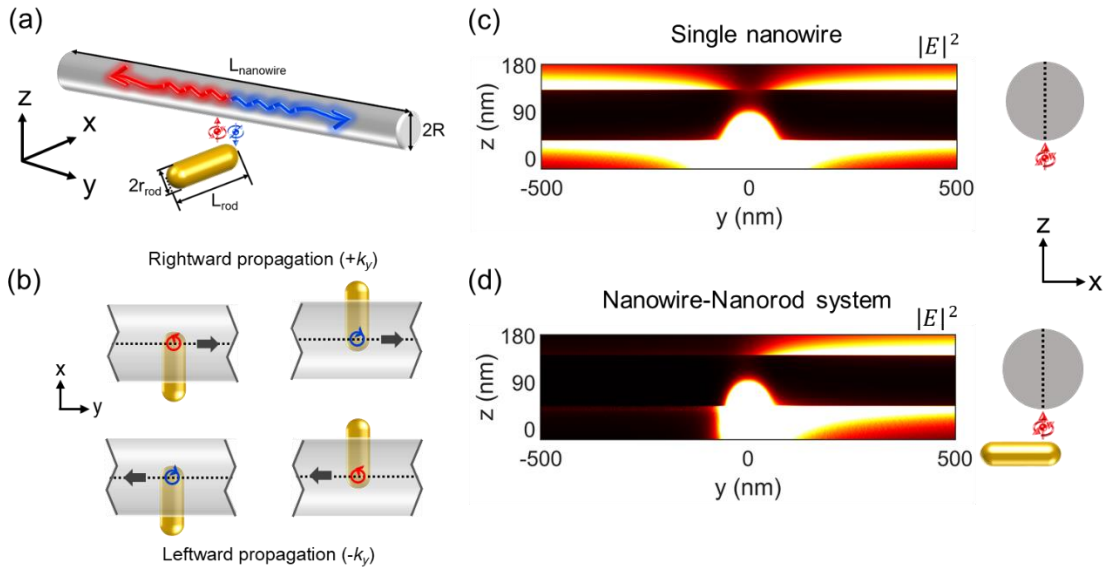

Fig. 1. (a) Model schematic of Nanowire-Nanorod system. Red dot: right-handed circular polarized (RCP) dipoles. Blue dot: left-handed circular polarized (LCP) dipoles. (b) Illustration of the handedness of the optical spin at the nanogap's center, which depends on the structure's configuration and propagation direction. Black arrow: propagation direction of waveguide mode. The electric field distributions of (c) pure nanowire and (d) Nanowire-Nanorod system are excited by RCP dipoles. The schematics on the right panel of (c) and (d) correspond to the structures on the left panel.

## 2. Results and discussion

Let us consider a gap plasmon structure consisting of a silver nanowire and a gold nanorod, as shown in Fig. 1(a). A gap of 10 nm is formed between the silver nanowire and the gold nanorod. Within a highly confined nanogap near-field, a TSAM perpendicular to the propagation direction naturally appears due to the spin-orbit interaction of light. That provides a one-to-one relationship between the handedness of the optical spin and the propagation direction of the photonic mode, so-called spin-momentum locking, resulting from their time-reversal symmetry 38. Therefore, the information on spin angular momentum can be directly transferred to the propagation direction of light. Due to the chiral coupling between the nanoscale light sources and the local direction-locked transverse optical spin, the emission with opposite chirality will couple with the plasmonic eigenmodes propagating in the opposite directions, as shown in Fig. 1(b). Moreover, for different hybrid system configurations, the sign of handedness of transverse spin at the center spot of the nanogap is opposite, even if the propagation direction of the mode is the same. Fig. 1(b) qualitatively illustrates that a combination of the hybrid system configuration and waveguide mode's propagation direction determines the handedness of transverse optical spin from a right/left-propagating mode. We simulated the electric field distribution of the proposed Nanowire-Nanorod system using a three-dimensional finite-difference time-domain (FDTD) method. The nanowire is represented by cylinders, 100 nm in diameter and 7 um in length. The dimensions of the nanorod are 15 nm in radius and 130 nm in length. The refractive index of the silver nanowire is $\varepsilon_{Ag}$ = -22.01 +0.41i at $\lambda$ = 686 nm from Johnson and Christy 39. The refractive index of the environment is 1. The chiral light source is simulated by placing two orthogonal linearly polarized dipole sources with a phase difference of 90 degrees 22. Fig. 1(c) and Fig. 1(d) depict electric field distribution when the electromagnetic mode is excited by the RCP dipoles with $\lambda$ = 686 nm. As shown in Fig. 1(c), when the RCP emitter is located at the center of the pure nanowire, the intensities of the guided modes coupled to both ends of the nanowire are equal due to the in-plane mirror symmetry. However, the emission direction becomes different when the emitter is in the nanogap between the nanorod and nanowire. In Fig. 1(d), the guided mode of the hybrid structure has the property of unidirectional transmission. Asymmetric directional transmission is the main feature of coupling circular dipoles to gap plasmon systems. We place a power monitor at each end of the nanowire to record the emission intensity from the chiral light source coupled to the two ends of the nanowire through the gap plasmons 22. The directionality is calculated from the intensity of light transmitted to both ends of the nanowire, $D = (T_L - T_R)/(T_L + T_R)$, where $T_{L/R}$ is the amount of light transmitted to the left/right end.

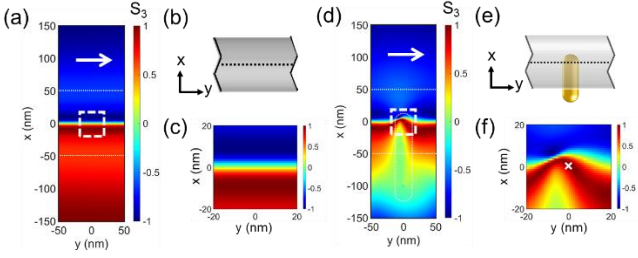

Fig. 2. (a) Density of transverse optical spin of the pure nanowire (b). (d) Density of transverse optical spin of Nanowire-Nanorod system (e). White dotted lines highlight the outlines of nanowire and nanorod. (c) and (f) are enlarged views of boxes outlined by white dashed lines in (a) and (d), respectively. Cross: position at x=0, y=0. White arrow: the propagation direction of the plane wave as the excitation source.

To quantify the magnitude of the transverse optical spin near the silver nanowire, we calculate the density of the transverse optical spin of pure nanowire and the gap plasmon system as a function of position in the x-y plane, which is the oscillating plane of the dipoles (in the middle of the gap), corresponding to the stokes parameter, $S_3 = -2\text{Im}(E_x E_y^*)/(E_x^2 + E_y^2)$. The TSAM is obtained by setting a right/left-propagating wave at the left/right end of a nanowire with a TE plane wave as the excitation source. Unidirectional transmission can be achieved via circularly polarized dipoles coupled to the evanescent wave with the corresponding TSAM. When the RCP dipoles are coupled to the guided mode with a positive sign of TSAM, the direction of photon transmission is also determined due to spin-direction locking. As shown in Fig. 2(a), the TSAM at the center of the nanowire (x = 0, y = 0) is zero, and the signs of TSAM on both sides of the nanowire are opposite, as expected 22. As a result, the amount of light coupled to both ends of the nanowire from nanoscale light sources at the central position is equal, as shown in Fig. 1(c), while the emission of RCP dipoles in the -x half-space couples to the right-propagating mode (see Fig. 3(b)). For the hybrid system in Fig. 2(e), at the central position of the nanogap (white cross), the guided mode has localized transverse spin density near unity with a positive sign, as shown in Fig. 2(f). The positive TSAM of the guided mode can be excited by the centrally located RCP dipoles, so the emission is coupled to the corresponding mode, the right-propagating mode, as shown in Fig. 1(d). When the propagation direction of the mode reverses, so is the helicity sign in all places (See Fig. S1). To further support that the directional emission is indeed caused by chiral interaction between the chiral dipoles and the TSAM of waveguide mode, the directionality as a function of the position x and y of the RCP dipoles is shown in Fig. S2 and Fig. S3, respectively. Interestingly, opposite handedness can be found on the same side of the nanowire for the gap plasmon mode. Notably, it occurs in a region 80 nm away from the center of the wire, where the mode's amplitude and its effect on the local density of optical states are not negligible. To facilitate the study of the factors affecting the directional coupling, we fix the circular dipoles at the nanowire's centrosymmetric positions (x = 0, y = 0). The coupling direction of the nanoscale light sources will depend on the handedness of the localized optical spin and the propagation direction of the guided mode of the hybrid structure.

To explore the effect of the structural symmetry of the system on the coupling directionality of chiral nanoscale light sources, we investigate the directionality of the RCP emitter as a function of $x_c$ (i.e., the distance between the nanorod and nanowire) as shown in Fig. 3(a). The position of the nanorod relative to the nanowire determines the configuration of the hybrid system, such as axisymmetric or non-axisymmetric configuration. Numerical results show that, for negative $x_c$, the emission of the circular dipoles is coupled to the eigenstate propagating from left to right. While for positive $x_c$, the emission direction is reversed. When the chirality of circular dipoles reverses, the propagation direction is also opposite. For the hybrid system with non-axisymmetric configurations, $x_c \neq 0$ and the directionality $D$ reach more than 75%. The highest directionality $D_{max}$ is 94.3% when $x_c = \pm 5$ nm. Due to inversion symmetry, when the positions of the nanorod are axisymmetric about the axis of the nanowire, the directional coupling is in the opposite direction, i.e., $D_x = -D_{x'}$, where x = -x'. Therefore, when the light source position is fixed in the proposed structure, the emission direction of the nanoscale light source can be tailored by tuning the positions of the nanorod relative to the nanowire. In Fig. 3(b), we show the directional coupling of circular dipoles to a nanowire as a function of the position of dipoles $x_e$. Compared with the proposed hybrid system, the coupling directionality is zero as the circular dipoles are located at the pure nanowire's center, i.e., $x_e = 0$.

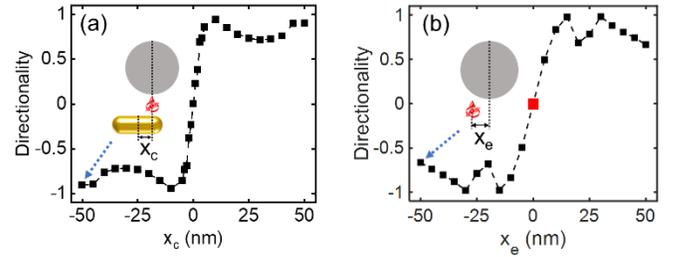

Fig. 3. Comparison of directional emission between (a) hybrid structure and (b) pure nanowire.

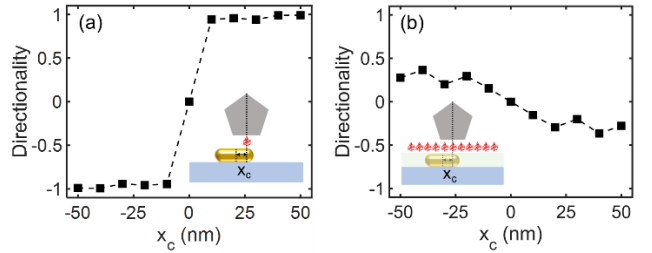

Fig. 4. The directional emission of (a) a single chiral light source and (b) multiple chiral light sources for the hybrid structure consisting of a pentagonal silver nanowire and a gold nanorod placed on the glass substrate.

In experimental preparation, nanowires usually have different cross-sections due to crystal structure constraints, which also change the property of chiral coupling. We replace the circular cross-section nanowire with a pentagonal nanowire (Fig. 4). The radius of the pentagonal nanowire is 50 nm. The simulation environment for the above directional calculation results is an ideal vacuum medium. Considering a more feasible dielectric environment, the hybrid structure is placed on a glass substrate. In this configuration, Fig. 4(a) shows the highest directionality of a chiral light source, near 99%. The luminescence of two-dimensional materials could be seen as the superposition of countless circular dipoles. Next, the case of placing 33 circular dipoles in the gap is calculated to represent the manipulation of the hybrid structure on bulk materials such as two-dimensional materials, as shown in Fig. 4(b). When the hot spot area is filled with chiral light sources, it can be seen that the directionality approaches 37%, which is lower than that of a single chiral light source.

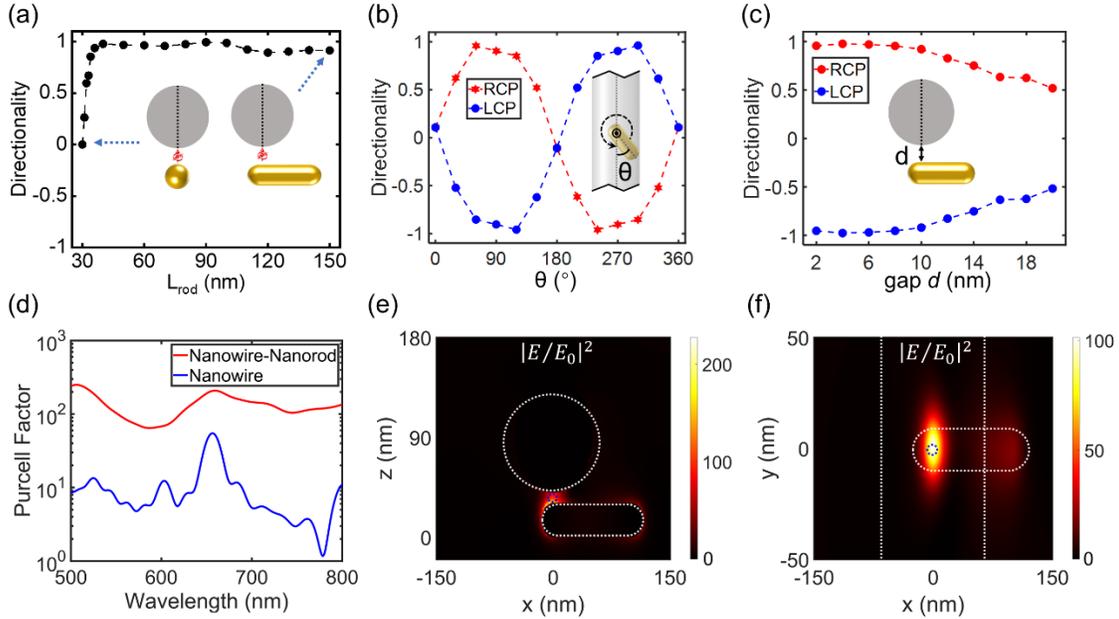

Fig. 5. The emission directionality is dependent on (a) nanorod aspect ratio, (b) orientation of nanorod, and (c) gap width. (d) Comparison of Purcell factor between the proposed hybrid system and pure silver nanowire. (e)-(f) Cross-sectional electric field intensity distribution of the gap plasmon mode in the x-z and x-y planes. Blue circle: emitter.

Next, we study the effect of the nanorod aspect ratio, the orientation of the nanorod, and the gap width on the corresponding coupling direction. In Fig. 5(a), we focus on the dependence of directionality on the nanorod aspect ratio. When the aspect ratio of the nanorod is 1, the structure is asymmetrically composed of nanosphere and nanowire. Thus, the emission from the circular dipoles coupled to both ends of the nanowire is equal. For a gap width of 10 nm, the coupling directionalities reach 92%, with an aspect ratio of 1.33 to 5. The maximum directionality is 99.5% when the aspect ratio is 3. The hybrid structure with nanorods supporting different LSPR wavelengths can achieve unidirectional emission, so the system has a broad working bandwidth. Fig. 5(b) shows the coupling directionality as a function of the angle θ between the longitudinal axes of the nanorod and nanowire. Since the orientation also affects the symmetry, we can see that the directionality changes periodically as the angle from 0 to 360°. By optimizing the orientation, a maximum directionality of 96.1% is obtained. For θ = 0° or 180°, the calculated directionality is not equal to 0, resulting from the reflection at the ends of the wire. Further, we investigate the effect of the gap width $d$ on the directional coupling, as shown in Fig. 5(c). In the 2 to 8 nm range, the directionalities reach 95%. As $d$ increases, the directionality gradually decreases to 62.4% at $d = 18$ nm. Since the gap width determines the local confinement of the gap plasmon mode and thus affects the directional coupling, the hybrid structure provides robust directionality remaining higher than 50% in the wide gap range of 2 to 20 nm. The comparison of Purcell enhancement between the hybrid system and pure nanowire is shown in Fig. 5(d). The circular dipoles confined in the nanogap of the proposed system obtained a ~159-fold Purcell enhancement, which is 15 times larger than that around pure nanowire. Fig. 5(e, f) shows the calculated cross-sectional electric field distributions of the gap plasmon modes in the x-z and x-y planes under the excitation of the circularly polarized wave by setting two orthogonal linearly polarized plane wave sources with a phase difference of 90 degrees. Within the 10-nm-wide gap between nanorods and nanowires, the localized electric field intensity enhancement by two orders of magnitude provides high excitation and emission rates for nanoscale light sources.

Finally, we discuss the experimental possibilities of the present scheme. Silver nanowires and gold nanorods are widely used in nanophotonic structures and have mature preparation techniques, such as chemical synthesis methods 40-41. The chiral light sources can be 2D TMDCs. Chiral emission can be achieved since the circular polarization state of the emitted photons of the TMDCs can be changed by switching the polarization state of the incident light, following the so-called valley-dependent optical selection rule 42. The system may be assembled in three steps. First, the nanorods are dispersed and deposited on the substrate. Then, 2D TMDC and silver nanowires could be sequentially assembled onto the substrate by dry transfer 43. Further, the relative position between the nanorod and nanowire could be tuned by nanomanipulation technology to adjust the configuration 33. In practical applications, the ability to steer the propagation direction needs to be considered. By assembling two nanorods on either side of the nanowire, the gap plasmon modes with different propagating directions can be selectively excited by changing the excitation position. Alternatively, switching the incident polarization of the excitation can lead to emission with opposite chirality, which will couple with the waveguide mode in the opposite direction. Furthermore, selecting emitters with different emission bands can also enable different coupling directions (see Fig. S4).

## 2. Conclusion

In conclusion, we demonstrate that a gap plasmon system consisting of gold nanorods and silver nanowires can control the directional coupling of chiral nanoscale light sources. The emission direction of chiral nanoscale light sources can be optimized by changing the configuration, such as the nanorod's aspect ratio, the nanorod's position and orientation relative to the nanowire, and the width of the gap. The optimal coupling directionality can reach 99.5% by optimizing the structural parameters. We also found that for a light source with the same handedness, if the position or orientation of the nanorod is symmetric about the axis of the nanowire, the emission will couple to guided mode propagating in the opposite direction. Additionally, the localized field enhancement within the nanogap reaches a factor of ~1500, thereby enhancing the excitation and emission rates of light sources. The present scheme can modify the emission directionality of valley excitons and corresponding photonic paths of valleytronics, which provides a new method for transmitting valley information and manipulating the chirality degree of freedom in on-chip photonic circuits. Besides, the gap-plasmon chiral coupling system has potential applications in chiral quantum networks, quantum gates, and quantum bits routers.

**Funding.** This work was supported by the National Key Research and Development Program of China (Grant No. 2022YFA1604304), the Guangdong Major Project of Basic and Applied Basic Research (Grant No. 2020B0301030009), and the National Natural Science Foundation of China (Grant Nos. 92250305).

**Disclosures.** The authors declare no conflicts of interest.

**Data availability statement.** The data cannot be made publicly available upon publication because no suitable repository exists for hosting data in this field of study. The data that support the findings of this study are available upon reasonable request from the authors.